\input harvmac.tex

\def\np#1#2#3{Nucl. Phys. {\bf B#1} (#2) #3}
\def\pl#1#2#3{Phys. Lett. {\bf #1B} (#2) #3}
\def\prl#1#2#3{Phys. Rev. Lett. {\bf #1} (#2) #3}
\def\prd#1#2#3{Phys. Rev. {\bf D#1} (#2) #3}

\def\jhep#1#2#3{JHEP {\bf#1}(#2) #3}
\def\jmp#1#2#3{J. Math Phys. {\bf #1} (#2) #3}

\def\IB{\relax\hbox{$\inbar\kern-.3em{\rm B}$}}
\def\IC{\relax\hbox{$\inbar\kern-.3em{\rm C}$}}
\def\ID{\relax\hbox{$\inbar\kern-.3em{\rm D}$}}
\def\IE{\relax\hbox{$\inbar\kern-.3em{\rm E}$}}
\def\IF{\relax\hbox{$\inbar\kern-.3em{\rm F}$}}
\def\IG{\relax\hbox{$\inbar\kern-.3em{\rm G}$}}
\def\IGa{\relax\hbox{${\rm I}\kern-.18em\Gamma$}}
\def\IH{\relax{\rm I\kern-.18em H}}
\def\IK{\relax{\rm I\kern-.18em K}}
\def\IL{\relax{\rm I\kern-.18em L}}
\def\IP{\relax{\rm I\kern-.18em P}}
\def\IR{\relax{\rm I\kern-.18em R}}
\def\IZ{\relax\ifmmode\mathchoice
{\hbox{\cmss Z\kern-.4em Z}}{\hbox{\cmss Z\kern-.4em Z}}
{\lower.9pt\hbox{\cmsss Z\kern-.4em Z}}
{\lower1.2pt\hbox{\cmsss Z\kern-.4em Z}}\else{\cmss Z\kern-.4em Z}\fi}

\def\II{\relax{\rm I\kern-.18em I}}

\chardef\tempcat=\the\catcode`\@
\catcode`\@=11
\def\cyracc{\def\u##1{\if \i##1\accent"24 i
    \else \accent"24 ##1\fi }}
\newfam\cyrfam
\font\tencyr=wncyr10
\def\cyr{\fam\cyrfam\tencyr\cyracc}


\def\CN {{\cal N}}

\def\CR {{\cal R}}




\def\Tr{{\rm Tr}}

\def\inbar{\,\vrule height1.5ex width.4pt depth0pt}
\font\cmss=cmss10 \font\cmsss=cmss10 at 7pt


\lref\juanads{J. Maldacena, hep-th/9712200}
\lref\ksorb{S. Kachru and E. Silverstein, hep-th/9802183}
\lref\dh{L.~Dixon and J.~Harvey, Nucl. Phys. {\bf B} 274 (1986) 93\semi
N.~Seiberg and E.~Witten, 
Nucl. Phys. {\bf B} 276 (1986) 272}

\lref\dmquiver{M.R. Douglas and G. Moore, hep-th/9603167}
\lref\dgmorb{M.R. Douglas, B.R. Greene and D.R. Morrison,
``Orbifold Resolution by D-branes'', \np{505}{1997}{84}; hep-th/9704151}
\lref\dougegs{M.R. Douglas, ``Enhanced Gauge Symmetry
in M(atrix) theory'', \jhep{007}{1997}{004}; hep-th/9612126}
\lref\jmorb{C.V. Johnson and R.C. Myers, ``Aspects of
Type IIB Theory on ALE Spaces'', \prd{55}{1997}{6382}, hep-th/9610140}
\lref\gipol{E.G. Gimon and J. Polchinski, ``Consistency Conditions for
Orientifolds and D Manifolds'', \prd{54}{1996}{1667}, hep-th/9601038}
\lref\polten{J. Polchinski, ``Tensors from K3 Orientifolds'',
\prd{55}{1997}{6423};
hep-th/9606165}

\lref\kmvgeom{S. Katz, P. Mayr and C. Vafa, ``Mirror Symmetry and Exact
Solution of 4D $\CN=2$ Gauge Theories -- I'', hep-th/9706110}
\lref\branegeom{H. Ooguri and C. Vafa, ``Geometry of $\CN=1$ dualities
in four dimensions'', \np{500}{1997}{62}; hep-th/9702180}

\lref\kleb{S. Gubser and I. Klebanov,``Absorption by Branes
and Schwinger Terms in the World Volume Theory,'' Phys. Lett. {\bf  
B413} (1997)
41.}

\lref\mfour{E. Witten, ``Solutions of Four-dimensional Field Theories
via M-theory'', \np{500}{1997}{3}; hep-th/9703166}

\lref\aspegs{P.S. Aspinwall, ``Enhanced Gauge Symmetries and K3  
Surfaces'',
\pl{357}{1995}{329}; hep-th/9507012}
\lref\bsvegs{M. Bershadsky, V. Sadov and C. Vafa, ``D-strings on  
D-manifolds'',
\np{463}{1996}{398}}

\lref\kwdiscrete{L.M. Krauss and F. Wilczek, ``Discrete Gauge Symmetries
in Continuum Theories'', \prl{62}{1989}{1221}}

\lref\sixteen{N. Seiberg, ``Notes on Theories with 16 Supercharges'',
hep-th/9705117}

\lref\reidrev{M. Reid, ``McKay correspondence'', alg-geom/9702016}

\lref\suthree{W.M. Fairbanks, T. Fulton and W.H. Klink, ``Finite and
Disconnected Subgroups of $SU(3)$ and their Application to the
Elementary Particle Spectrum'', \jmp{5}{1964}{1038}}
\lref\sosix{W.~Plesken and M.~Pohst, Math. Comp. {\bf 31} (1977) 552}
\lref\cave{A.M.~Polyakov, hep-th/9711002, hep-th/9809057}
\lref\ktone{I.~Klebanov and A.~Tseytlin, hep-th/9811035, hep-th/9812089}
\lref\kttwo{I.~Klebanov and A.~Tseytlin, hep-th/9901101}
\lref\lnv{A.~Lawrence, N.~Nekrasov and C.~Vafa,  hep-th/9803015}
\lref\bkv{M.~Bershadsky, Z.~Kakushadze and  C.~Vafa, hep-th/9803076}
\lref\bj{M.~Bershadsky and A.~Johansen, hep-th/9803249}
\lref\bg{O.~Bergman and M.~Gaberdiel, hep-th/9701137}
\lref\minahan{J.~Minahan, hep-th/9811156, hep-th/9902074}
\lref\tz{A.~Tseytlin and K.~Zarembo, hep-th/9902095}
\Title{ \vbox{\baselineskip12pt\hbox{hep-th/9902110}
\hbox{HUTP-99/A007}
\hbox{ITEP-TH-6/99}}}
{\vbox{
 \centerline{On Non-Supersymmetric  CFT in Four Dimensions}}}
\medskip
\centerline{Nikita Nekrasov $^{1,2}$ and  Samson L. Shatashvili\footnote{$^{3}$}{On leave of 
absence from St. Petersburg Steklov Mathematical
Institute}}

\vskip 0.5cm

\centerline{\cyr $^{1}$ Institut Teoretiqesko\u\i\quad i E1ksperimentalp1no\u\i\quad Fiziki, 
117259, Moskva, Rossiya}

\centerline{$^{2}$ Lyman Laboratory of Physics,
Harvard University, Cambridge, MA 02138}
\centerline{$^{3}$ Department of Physics, Yale University, New Haven, CT
06520 USA}

\medskip
\bigskip
\noindent
We show that the $\CN=0$ theories on the self-dual $D3$-branes of Type 0
string theory are in the class of the previously considered
{\it tadpole-free  orbifolds} of $\CN = 4$ theory (although they have
$SO(6)$ global symmetry) and hence
have vanishing beta function in the planar limit to all  orders
in 't Hooft coupling. Also, all planar amplitudes in this theory are
equal to those of $\CN = 4$ theory, up to a rescaling of the coupling.

\medskip
\Date{February 1999}

\noindent
{\bf Vanishing of the planar beta function.}
Recently the non-supersymmetric gauge theories were studied
in the framework of string theory/gauge theory correspondence
\cave\ktone\kttwo\minahan. 
The 
case \juanads\ of $\CN=4$ super-Yang-Mills theory can be modified in several ways
leading to the theories with lower supersymmetry. One possibility is
to restrict the set of fields of $\CN=4$ theory according to
the following principle: take a discrete subgroup $\Gamma$ 
of the R-symmetry group $SO(6)$ and let it act on the color indices in some
representation $\CR$: $g \mapsto \gamma_{g} \in U ({\rm dim} {\CR})$.
Let $\Phi^{\alpha}$ denote a field of the $\CN=4$ theory.
The index $\alpha$ belongs to the representation of the $R$-symmetry
group, and both color and  Lorentz indices are omited. Impose the invariance 
condition:
\eqn\inv{
\Phi^{\alpha} =  g^{\alpha}_{\beta} \left[ \gamma_{g}^{-1} 
\Phi^{\beta} \gamma_{g}\right]
}
for any $g \in \Gamma$. It is known \bkv\bj\ that if the
representation
$\CR$ is such that for any group element $g \neq 1$
\eqn\reg{
{\Tr}_{\CR} \gamma_{g} = 0}
then the planar  graphs in the resulting theory are equal to
those of the parent $\CN=4$ theory up to a rescaling of coupling.
All these theories occur in the studies of the Type IIB theory
compactified on ${\IR}^{1,3} \times {\IR}^{6}/{\Gamma}$ \dmquiver\ and they
were suggested as examples of large $N$ conformal theories dual to 
certain string theory backgrounds \ksorb. They are known to be one-loop
finite for all $\Gamma$ \lnv\ while for higher loops they are known to
have vanishing gauge beta functions in the planar limit (see \lnv\ for
two loops in the $\CN=1$ case, \bkv\bj\ for general case).

I.~Klebanov and A.~Tseytlin \ktone\kttwo\
suggested to study the $D$-branes in the Type 0B theory \dh.
This theory has in the low-energy spectrum no fermions, the same NS bosons
as Type IIB theory but the doubled set of RR bosons. 
The doubled set of RR fields leads to the doubled set of D-branes. In
particular, there are two types of D3-branes: electric and magnetic
ones. It turns out that the low-energy theory on the stack of $N$
D3-branes is a truncated version of the $\CN =4$ theory. In the case of
pure electric or pure magnetic D3-branes one has a theory
of $U(N)$ gluons coupled to six adjoint scalars while in the case of
$(1,1)$ branes one has $U(N) \times U(N)$ gauge group, 
six adjoint scalars for each gauge factor and two sets of bi-fundamental
fermions $(N, \bar N)$ and $(\bar N, N)$. Those fermions correspond to the 
open strings connecting electric and magnetic branes \bg\ktone. 
In the paper \kttwo\
the two-loop finiteness of this self-dual theory was proven in the leading-$N$
approximation while \tz\ studied further aspects of self-dual threebranes.

The purpose of our paper is to point out that this theory has in fact
vanishing
planar beta function to all orders in 't Hooft coupling, just like
any other {\it orbifold} theory studied in \ksorb\lnv\bkv\bj. 

The point is that both the electric and self-dual theories are
also {\it orbifolds} of $\CN = 4$ theory. The slightly subtle point
is that the R-symmetry group of $\CN = 4$ theory is the spin cover of
$SO(6)$, i.e. $SU(4)$. This group has a center $Z \approx \IZ_{4}$.
The group $\Gamma = \IZ_{2} \subset Z$ can also be
 used for orbifolding just like any other subgroup 
of $SU(4)$. 

By inserting projectors on the $\Gamma$-invariant fields \inv\
into the 't Hooft diagrams one immediately sees that if the
representation of $\Gamma$ in the gauge group obeys \reg\ then the
planar gauge coupling beta function vanishes. 

For the self-dual theory one starts with $U(2N)$ $\CN = 4$ theory and
represents
$\Gamma = \{ 1, \omega \}$ in the $U(2N)$ as follows: 
$$
\gamma_{\omega} = \pmatrix{ 1 & 0 \cr 0 & - 1}
$$
where the blocks are $N \times N$. Clearly this representation
obeys \reg. 

For the purely electric theory one starts with 
$U(N)$ $\CN=4$ gauge theory and takes the trivial representation of
$\Gamma$ which does not obey \reg. Hence the theory has non-trivial
beta function even in the large $N$ limit.

\noindent
{\bf Another $SO(6)$ invariant theory}. 
Another $\CN=0$ theory with $SO(6)$ global symmetry is the one
corresponding to $\Gamma = \IZ_{4}$. We would start
with $U(4N)$ $\CN=4$ gauge theory and represent $\Gamma = \{ 1, \omega,
\omega^2,
\omega^3 \}$ as $\gamma_{g} = {\rm diag} \left( 1, g, g^2, g^3 \right) 
\otimes 1_{N \times N}$. In this case
the orbifold group acts on the scalars by changing the sign of all six of them.
The field content is\foot{We thank I.~Klebanov
and L.~Paniak for correcting
our error in the previous version of the paper}:
gauge group $U(N)^4$, two sextets of  bi-fundamental scalars $(N_{i}, \bar N_{i+2})$, 
four sextets of bi-fundamental fermions: $(N_i, \bar N_{i+1})$, $i = 0,1,2,3$,
$4 \equiv 0$.

\lref\kns{I.~Klebanov, N.~Nekrasov and S.~Shatashvili, to appear}

This theory has a string theory realization, as
Type IIB on ${\IR}^6/{\IZ}_{2}$, clearly
has vanishing planar beta function and hence would have dilaton-free
dual $AdS_5 \times {\bf RP}^5 $ background. It has interesting features 
which will be explored elsewhere \kns.

\noindent
{\bf Acknowledgements.} We are grateful to I.~Klebanov for discussions. 
The 
research  of N.~N.~ is supported by Harvard Society of Fellows, 
partly by NSF under
grant PHY-98-02-709, partly by {\cyr RFFI}
 under grant 98-01-00327 and 
by grant 96-15-96455 for scientific schools; that 
of S.~S. is supported by DOE grant
DE-FG02-92ER40704, by NSF CAREER award, by OJI award from DOE and by
Alfred P.~Sloan foundation.  

\listrefs
\bye